%% file: main.tex
\documentclass[acmtog,nonacm]{acmart}

\renewcommand\footnotetextcopyrightpermission[1]{}
\usepackage{booktabs} %
\usepackage{algorithmic}
\usepackage{wrapfig}

\usepackage[ruled]{algorithm2e} %

\SetAlFnt{\small}
\SetAlCapFnt{\small}
\SetAlCapNameFnt{\small}
\SetAlCapHSkip{0pt}

\makeatletter
\let\@authorsaddresses\@empty
\makeatother

\begin{document}
\title{Temporal Brightness Management for Immersive Content}

\author{Luca Surace}
\affiliation{%
  \institution{Università della Svizzera italiana}
  \country{Switzerland}}
\author{Jorge Condor}
\affiliation{%
  \institution{Università della Svizzera italiana}
  \country{Switzerland}
}
\author{Piotr Didyk}
\affiliation{%
 \institution{Università della Svizzera italiana}
 \country{Switzerland}}

\begin{teaserfigure}
 \includegraphics[width=\linewidth]{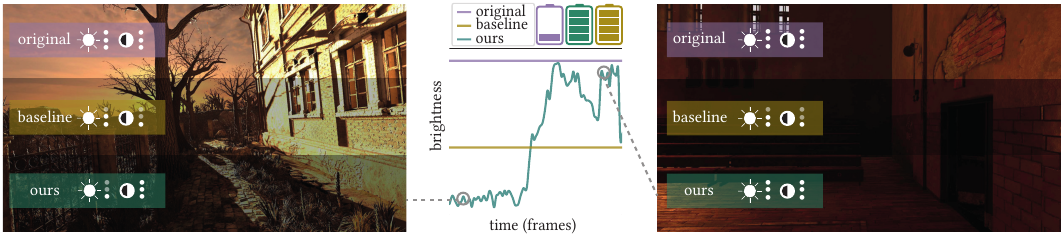}
 \centering
\caption{Our method adjusts the screen brightness based on the content of each frame. In the outdoor scene on the left, characterized by high brightness, our technique darkens the frame to a greater extent compared to a constant reduction (baseline) without significant losses in the visibility of image details. Conversely, in the darker indoor space shown on the right, where dimming the screen can easily lead to loss of visible image details, our technique increases the brightness to better preserve contrast. Throughout the entire sequence, our technique utilizes the same power budget as the baseline. }
\label{fig:teaser}
\end{teaserfigure}

\input{0-abstract}

\maketitle

\input{samplebody-journals}

\end{document}

%% file: 0-abstract.tex
\begin{abstract}

Modern virtual reality headsets demand significant computational resources to render high-resolution content in real-time. Therefore, prioritizing power efficiency becomes crucial, particularly for portable versions reliant on batteries. 
A significant portion of the energy consumed by these systems is attributed to their displays. Dimming the screen can save a considerable amount of energy; however, it may also result in a loss of visible details and contrast in the displayed content. 
While contrast may be partially restored by applying post-processing contrast enhancement steps, our work is orthogonal to these approaches, and focuses on optimal temporal modulation of screen brightness. 
We propose a technique that modulates brightness over time while minimizing the potential loss of visible details and avoiding noticeable temporal instability.
Given a predetermined power budget and a video sequence, we achieve this by measuring contrast loss through band decomposition of the luminance image and optimizing the brightness level of each frame offline to ensure uniform temporal contrast loss.
We evaluate our method through a series of subjective experiments and an ablation study, on a variety of content. We showcase its power-saving capabilities in practice using a built-in hardware proxy. Finally, we present an online version of our approach which further emphasizes the potential for low level vision models to be leveraged in power saving settings to preserve content quality.

\end{abstract}

%% file: samplebody-journals.tex
\input{1-introduction}

\input{2-related-work}

\input{3-method}
\input{4-results}

\input{5-power}

\input{6-real-time}

\input{7-limitations}

\input{8-conclusion}

\bibliographystyle{ACM-Reference-Format}
\bibliography{bibliography}

\input{9-figures}

%% file: 1-introduction.tex
\section{Introduction}
Power consumption is a fundamental axis of hardware design, particularly in edge and portable devices. Battery life and computational power form a careful balance, and efficient use of power resources can directly contribute to increased computational power budget for more realistic and interactive applications.

In virtual reality (VR) devices, dimming the screen is a common approach to energy preservation, but it encompasses a trade-off between power consumption and image quality. Recent work has highlighted this problem \cite{chen2024pea}, comparing various power-saving techniques and their respective impacts on perceived quality. Their analysis shows that, on OLED architectures with a 30\% power-saving target, uniform dimming ranks as the second-best, in terms of perceptual impact measured by Just-Objectionable Differences (JOD). Furthermore, when the power-saving target exceeds 60\%, uniform dimming becomes the most effective approach for maintaining visual quality. For more common Liquid Crystal Displays (LCDs), uniform dimming is the only method capable of reducing power consumption by 50\% without incurring a significant perceptual quality loss. %
Therefore, refining global dimming techniques is a promising avenue for advancing power-efficient display designs across various architectures. However, substantial corrections to brightness levels risk reducing content visibility and luminance contrast. 

Previous works address this problem with frame-dependent brightness scaling techniques, evaluated using standard image quality metrics \cite{pagliari2019low}. Additionally, user studies have assessed the acceptability of brightness degradation \cite{park2015saving}. Integrating post-processing steps with power-saving measures, including contrast enhancement \cite{pagliari2018lapse} and luminance retargeting \cite{wanat2014simulating} is a relatively simple approach to partly mitigate the degradation in quality. 
Unlike previous techniques, however, our method explicitly models luminance adaptation, resulting in imperceptible brightness modulations within VR environments.

In this paper, we present a novel technique for dynamically adjusting screen brightness to reduce display power consumption. Our method is formulated as an optimization problem based on a low-level human vision model that accounts for perceived luminance contrast. The optimization process aims to achieve a balance among three objectives. First, it aims to maintain power consumption at a specified input level. Second, it seeks to ensure consistent luminance contrast loss throughout the entire sequence. Finally, it guarantees that any variations in brightness over time are not perceived as artifacts, such as visible flickering. Here, we specifically design our methods for virtual reality displays, which allows us to leverage the fact that users immersed in a virtual environment without access to real-world stimuli exhibit low sensitivity to global luminance changes. We thoroughly validate our technique through subjective human experiments and measure its power savings on actual hardware. We demonstrate that our method not only improves perceived quality compared to a baseline solution which reduces the brightness uniformly over the sequence duration, but can also lead to improved task performance. Apart from the offline optimization, we also propose a simple online solution which realizes the same goals.

%% file: 2-related-work.tex
\begin{figure*}[t!]
    \centering
    \includegraphics[width=\textwidth]{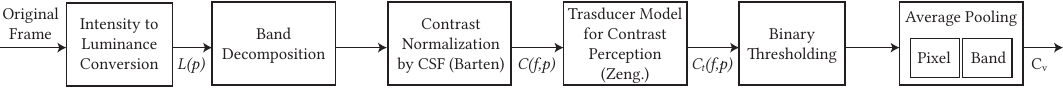}
    \caption{Pipeline of our method. The conversion to luminance is achieved using the luminance curve of the display and applying the display model from Mantiuk et al. (Equation \ref{eq:display-model}). The self-parameters for masking remain consistent with the original model proposed by Zeng et al. \shortcite{zeng2000point}: $\alpha = 0.7$ and $\beta = 0.2$. The output value $C_v$ represents the visible contrast in the frame.}
    \label{fig:pipeline}
\end{figure*}

\section{Related work}
\label{related-work}
Power reduction strategies vary depending on display technology. LCDs depend on a backlight filtered by a color matrix, making backlight dimming the main method for reducing power consumption. Conversely, OLED panels are self-emissive, allowing more precise control since the intensity and color of each pixel impact overall power use. Irrespective of the technology, reducing power can result in a perceived loss of image quality. Advanced techniques aim to balance power efficiency with image quality. Recently, \cite{chen2024pea} established a standard framework for quantifying the perceptual impacts and power savings of different solutions, which highlights global dimming as one of the most promising avenues for imperceptible power use reduction.

\paragraph{Display dimming}
The most dominant method for reducing power consumption is display dimming \cite{cheng2004power, choi2002low, gatti2002low, chen2024pea, park2015saving}. In addition to various models of power consumption and dimming strategies, some approaches focused on explicitly addressing the trade-off between power consumption and quality loss by using simple quality metrics like PSNR or SSIM as constraints \cite{kang2015PSNR, kang2015perceptual,chang2016}. More advanced methods integrate quality metrics as objectives within optimization processes. For instance, \citeauthor{pagliari2019low} \shortcite{pagliari2019low} employed a simplified SSIM metric in an optimization framework that determines the appropriate brightness for each displayed frame. In the context of immersive wide-field-of-view display, it is possible to apply brightness modulation only in the periphery, where it is less noticeable \cite{kim2020}. These simple metrics, however, are not capable of capturing the full complexity of the human visual system (HVS) and can lead to temporal artifacts or smaller potential power savings \cite{pagliari2019low}.

\paragraph{Color modulation}
In non-quantum dot self-emissive displays, like OLED panels, it is possible to enhance power efficiency by remapping pixel colors based on the efficiencies of primary colors. In certain cases, changing the color scheme or profile can be a viable method to save energy \cite{dong2011power,dong2011chameleon}. However, when it comes to natural content, remapping colors may produce undesirable effects. It has been demonstrated that this strategy can still be effectively applied in wide-field-of-view displays by taking advantage of the limitations of the HVS in peripheral vision and modulating colors only in those regions \cite{duinkharjav2022color}.

\paragraph{Compensation methods}
Dimming the display can significantly impact the quality of image details, color, and depth perception. A number of techniques have been dedicated to compensate for such undesired changes. An example of a simple compensation strategy is an increase in image brightness prior to display, which preserves the visibility of image details in dark image regions but introduces clipping to bright regions. Kerofsky and Dally \shortcite{kerofsky2006brightness} proposed to counteract this effect by smoothly attenuating bright image values to preserve bright image details. Such an approach is similar to other methods that apply global contrast adjustment to correct for the lost details \cite{choi2002low, cheng2004power,cho2013image}. The loss of spatial details can be more effectively addressed by local contrast enhancement methods. Such methods have been extensively studied in the context of tone mapping, with examples such as, \cite{mantiuk2008display,krawczyk2007contrast,eilertsen2016real}. They usually rely on accurate modeling of luminance perception and frameworks for local contrast processing. In the context of power-efficient displays, existing works proposed retargeting methods for matching image appearance between dark and bright conditions \cite{wanat2014simulating, ashraf2022suprathreshold}. It is important to note that while the visibility of image details is the primary visual cue affected by reduced brightness, others that rely on luminance patterns can also be inhibited. An example of such a cue is stereopsis \cite{didyk2012}. The problem was investigated in \cite{wolski2022dark}, where the authors proposed a method for compensating side effects of reduced luminance conditions on depth perception. Another issue related to brightness modulation can be temporal coherence. Rapid changes in luminance can lead to flickering artifacts, while slower modulation rates may result in brightness incoherence over time. Closest to our method are the techniques that focus on preserving temporal brightness coherency in the context of tone mapping \cite{boitard2012temporal, boitard2014zonal}. It is important to note that any post-processing effort to recover lost contrast is orthogonal to our work, and could be integrated into our optimization procedure to further reduce power consumption.

\paragraph{Discussion}
The method presented in this paper aims to enhance power efficiency through display dimming, aligning with existing methods in this category. However, there are significant differences between our approach and previous techniques. First, our method explicitly models luminance adaptation, which is crucial for understanding the reduction in perceived contrast due to decreased image brightness. Second, we rely on a precise perceptual model of perceived luminance contrast, in contrast to some general-purpose quality metrics that do not provide acccurate information about visibility of luminance contrast. Finally, our method focuses on brightness modulation that remains imperceptible to users. While many previous techniques were designed for general screens, such as cell phones, our approach specifically targets VR devices. This allows us to leverage the fact that, while immersed in a virtual environment without access to real-world stimuli, users exhibit very low sensitivity to gradual brightness changes. Through careful calibration of our method, we ensure that the brightness modulation remains imperceptible.

%% file: 3-method.tex
\section{Method}
\label{method}
When reducing the screen brightness, the visibility of luminance contrast in the image changes. The effect can be directly attributed to the reduction of the sensitivity of the HVS to luminance contrast when adaptation luminance decreases \cite{wanat2014simulating}. Consequently, screen dimming may result in a loss of visible luminance contrast, leading to poorer visibility of image details. %
Our goal is to achieve a specific average power budget for the display by intelligently dimming the screen based on the content being shown, preserving visible contrast. Since such temporal brightness modulation may lead to visible temporal artifacts, we constrain the modulation rate to remain smooth. Our method quantifies the loss of visible contrast by analyzing perceived luminance contrast modeled using multi-band frequency decomposition (Section~\ref{perceived-contrast}).In addition to quantifying the magnitude of contrast loss, our method also identifies the portions of the image where details become invisible due to the brightness reduction (Section~\ref{sec:contrast-loss}). An overview of this pipeline is shown in Figure \ref{fig:pipeline}. We use this per-frame prediction to optimize the brightness modulation for an entire video sequence offline (Section~\ref{sec:optimization}). %

\subsection{Perceived Contrast}
\label{perceived-contrast}
In order to locally measure where contrast is perceivable by a human observer, and quantify its magnitude per frame, we adapt \citeauthor{tursun2019luminance}~\shortcite{tursun2019luminance} approach. We analyze the contrast in a frame, as defined by \citeauthor{michelson1995studies}~\shortcite{michelson1995studies}, using a frequency band decomposition of the luminance image.
We first convert image intensity to luminance using \citeauthor{mantiuk2008display}~\shortcite{mantiuk2008display} display model:
\begin{equation}
    \label{eq:display-model}
    L(p) = (L')^\gamma \cdot (L_{max} - L_{black}) + L_{black} + L_{refl},
\end{equation}
where $L$ is the displayed luminance, $L'$ is the pixel value at location $p$, and $\gamma$ is the display gamma. $L_{max}$ denotes the peak display luminance, while $L_{black}$ is the display's black level. The term $L_{refl}$ refers to ambient light reflected from the display surface.
Next, we compute the Laplacian pyramid decomposition \cite{burt1983adelson} of the luminance image, obtaining the band-limited luminance difference, $\Delta L(f,p)$, at given frequencies $f$ and image locations $p$. To compute perceived contrast, we normalize the pyramid by the average luminance in the area $L_a(f,p)$, which is provided by the corresponding point in the Gaussian pyramid two levels down the resolution. Finally, contrast is normalized by the Contrast Sensitivity Function (CSF) \cite{barten2003formula}. The final perceived contrast value can be then computed as:
\begin{equation}
    C(f,p) = \frac{\Delta L(f,p)}{L_a(f,p) + \epsilon} \cdot S_{CSF} (f,L_a(f,p)),
\end{equation}
where the additional $\epsilon$ denotes a small value that prevents singularities in dark image regions. To incorporate the effect of the visual masking, we apply the transducer model of Zeng et al. \shortcite{zeng2000point} with their proposed default parameters and obtain the final perceived luminance contrast as
\begin{equation}
    C_t(f,p) = \frac{ \text{sign}(C(f,p)) \cdot \vert C(f,p) \vert ^{0.7}}
                    { 1 + \frac{1}{N} \sum_{q \in N(p)} \vert C(f,q) \vert ^ {0.2}},
\end{equation}
where $N(p)$ denotes a 5 x 5 neighborhood in the same band. Differently from the work of Tursun et al. \shortcite{tursun2019luminance}, we do not model retinal eccentricity, relying on a simpler model for the CSF, since we target common global-backlit panels. This may be of interest, however, in gaze-contingent scenarios using self-emitting displays like OLED or QD-LED, and panels with local area dimming.

\subsection{Contrast Loss}
\label{sec:contrast-loss}
\begin{figure}
    \centering
    \includegraphics[width=\columnwidth]{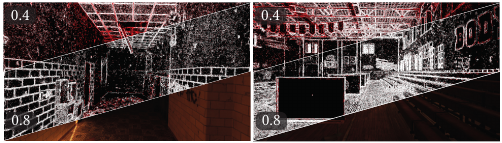}
    \caption{Visible contrast (white) at the highest frequency level of the band decomposition after binary thresholding, and regions where contrast loss occurs between the dimmed frame and the original frame (red), for target average brightness levels of 0.4 (top) and 0.8 (middle). The original frame is displayed at the bottom of the image. The lower the backlight brightness, the higher the amount of detail lost.}
    \label{fig:thresholds}
\end{figure}

We use the perceived luminance contrast measure to compute contrast loss due to screen dimming. To do so, we compare perceived contrast pyramids of the reference luminance image $I$ and its dimmed version $D=b \cdot I$, where $b$ is a dimming factor. We obtain the portion of the contrast pyramids containing visible contrast using

\begin{equation}
\label{eq:visiblecontrastpyramid}
    C_v = \frac{1}{N_l} \sum_{f} \sum_{p} \frac{\mathbb{I}\left(C_{t}(f,p)>1\right)}{N_p^f},
\end{equation}
where summation over $f$ iterates over pyramid levels, discarding the coarsest level containing the average value of the image; while 
the summation over $p$ iterates over all pixels in the level. Additionally, $N_l$ denotes the number of levels, $N_p^f$ is the number of pixels in the layer corresponding to frequency $f$, and $\mathbb{I}$ is the indicator function which returns $1$ if the condition given as the argument is true, and $0$ otherwise. Since we normalize by the CSF, we set the threshold value within the argument to 1, as it corresponds to the smallest contrast level that the HVS can reliably perceive, serving as a lower limit for visibility under ideal conditions.
Figure \ref{fig:thresholds} illustrates the effect of $\mathbb{I}$ on one contrast pyramid level. Finally, we compute the value of Equation~\ref{eq:visiblecontrastpyramid} both for the reference images as $C_v^I$ and for the dimmed version $C_v^D$ and define the contrast lost due to the dimming $I$ by a factor $b$ as
\begin{equation}
    \mathcal{L}_c(I,b) = \left| 1 - \frac{C_v^D}{C_v^I} \right|, \quad \text{where} \quad D = b \cdot I.
\end{equation}

\subsection{Optimization}
\label{sec:optimization}
Our objective is to modulate brightness such that it achieves a target power budget on average on a specific display, while minimizing contrast loss. Additionally, the rate at which brightness is modulated should be restricted to avoid temporal artifacts. 

We formulate this as an optimization problem. The input is a video sequence represented by a set of $N$ luminance images $\{I_i\}$ and a target power budget $\mathcal{P}_{base}$. 
The goal of the optimization is to find a set of per-frame brightness reduction factors $\{b_i\}$, fulfilling our design goals outlined above. More formally, we define the following optimization as:
\begin{equation}
\label{eq:contrast_minimization}
\begin{aligned}
& \underset{\{b_i\}}{\text{minimize}}
& & \sum_{i=1}^N \sum_{j=i+1}^N \left| \mathcal{L}_c(I_i,b_i) - \mathcal{L}_c(I_j,b_j) \right| \\
& \text{subject to}
& & \sum_{i=1}^{n-1} \left| \frac{dL_a(I_i,b_i)}{dt} - \Delta L \right| = 0\\
& & & \sum_{i=1}^N \frac{\mathcal{P}_i}{N} = \mathcal{P}_{base},\\
& & & \forall_{0<i\leq N} \quad 0<b_i<1, \\
\end{aligned}
\end{equation}
where $\frac{dL_a(I_i,b_i)}{dt}$ denotes the derivative of the average luminance of each frame and $\Delta L$ is the maximum undetectable rate of luminance change, which we derive in Section~\ref{sec:calibration}. Finally, $\mathcal{P}_i$ is the power consumption required for the display to show the frame $I_i$ with the dimming factor $b_i$, and it is dependent on the specific display. 

\subsection{Brightness modulation rate calibration}
\label{sec:calibration}
We find the maximum rate $\Delta L$ at which brightness is modulated according to human perception. We evaluate human sensitivity to global luminance changes by conducting a subjective two-alternative forced choice (2AFC) experiment. We then use the resulting data to ensure that our modulation does not lead to excessively steep transitions, minimizing the risk of producing visible artifacts. Each specific display requires its own calibration.

\begin{wrapfigure}{r}{0.5\columnwidth}
    \centering
    \includegraphics[width=0.48\columnwidth]{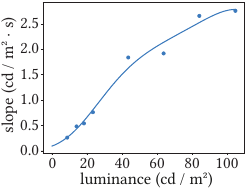}
    \caption{Correlation between luminance and the slope at the visibility threshold (probability of 75\% of detecting the change). The curve is a fourth-degree polynomial fitted to the data points.}
    \label{fig:slopes}
\end{wrapfigure}

In the experiment, 10 participants were presented with a white screen of constant brightness, which was linearly decreased over a five-second interval and followed by an additional second of constant brightness. Initial brightness varied from a predefined set of brightness factors $B = \{0.2, 0.4, 0.6, 0.8, 1.0\}$. Additionally, different slopes were tested for each value. Participants were instructed to indicate whether they perceived a change in brightness by responding with either "yes" or "no". The threshold for detecting brightness changes was determined as the value corresponding to a 75\% detection rate for each factor, calculated by fitting a logistic function to the answers. After determining the thresholds, we measured the displayed luminance corresponding to the initial and final brightness levels with a luminance meter. We repeated the procedure using a gray level to cover the relevant luminance range. In Figure \ref{fig:slopes} we show the curve obtained from the experimental data.

\subsection{Implementation}
We minimize the objective function in Eq. \ref{eq:contrast_minimization} using the Sequential Least-Squares method \cite{carayannis1983fast}. Additionally, we parameterize our display model (Equation \ref{eq:display-model}) setting ambient light reflected from the display surface to zero, which is a fair assumption on VR setups. We initialize the optimization process with a sequence of constant brightness factors corresponding to the desired target power reduction, perturbed by a small random deviation, $\delta$. Since $\mathcal{L}_c(I_i,b_i)$ is repeatedly evaluated, for each image $I_i$, we precomputed the term for a set of brightness factors $\{ 0.2, 0.4, 0.6, 0.8, 1.0 \}$ and linearly interpolate between the obtained values during the optimization for efficiency reasons. We ran all our optimizations until convergence with a tolerance value of $10^{-8}$. An optimization of 3000 frames takes around three minutes single-threaded on CPU (Intel Core i7-9700K @ 3.60GHz). 

%% file: 4-results.tex
\section{Results}
\label{sec:results}
\begin{figure}
    \centering
    \includegraphics[width=\columnwidth]{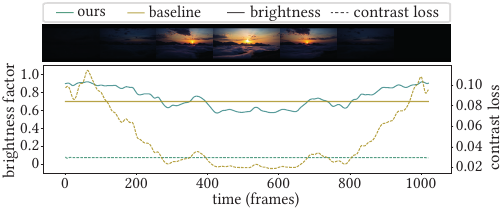}
    \caption{Time-lapse sequence capturing the Sun's movement (top, input frames). Our technique optimizes power allocation by spending more of the "brightness budget" during the initial and final parts of the video, where the frames are darker. At the same time, during the middle section, the dimming can be more aggressive.}
    \label{fig:sunrise}
\end{figure}

We test our method with different video sequences exhibiting a variety of scenery, both synthetic (Figures \ref{fig:teaser}, \ref{fig:brightness-contrast}) and real (Figure \ref{fig:sunrise}).

\begin{figure}
    \centering
    \includegraphics[width=\columnwidth]{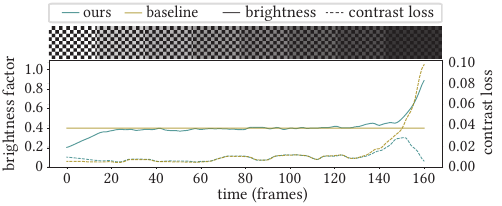}
    \caption{We apply our method to a sequence of checkerboard patterns (top, input frames), where the white squares gradually transition to black through various shades of grey. Our method compensates for the increasing loss of visible contrast by enhancing the brightness accordingly.}
    \label{fig:scacchiera}
\end{figure}

We visually showcase our algorithm in Figure \ref{fig:scacchiera}, applied to checkerboard patterns. Initially, the tiles are black and white, maximizing contrast, which allows the brightness factor to remain low. As the white tiles slowly approach black, the algorithm compensates for the loss of contrast by increasing the brightness factor. In a second, more realistic scenario, we run our optimization on a timelapse video sequence of a sunrise (Figure \ref{fig:sunrise}). The video is reversed at half of its duration. During the darker intervals, when the sun is not visible, brightness is high. Conversely, brightness decreases when the sun rises above the horizon.

In order to validate our approach, we ran an extensive set of experiments on human subjects to measure the impact of our modulation on perceived quality and contrast. For all our perceptual experiments, we used a Varjo XR-3 headset. We include details on its hardware capabilities on Supplementary materials.

\paragraph{Stimuli}
We created four different camera paths (\textsc{library}, \textsc{school}, \textsc{living room}, \textsc{basement}) from two high-quality 3D environments containing a variety of indoor, outdoor and differently illuminated spaces (Figure~\ref{fig:brightness-contrast}). We generated four video sequences from these paths, each targeting a different average brightness factor: $30\%$, $40\%$, $50\%$, and $70\%$ of the maximum brightness. We compute with our method (Figure \ref{fig:pipeline}) the optimal brightness modulation (referred to as \textit{brightness factor}) for the four sequences.
For comparison, we render a \textsc{baseline} alternative, which simply applies constant dimming to achieve the same target power consumption.

\subsection{Detail Preservation}
\label{sec:quality}
The first experiment aimed to assess the quality of perceived details in different environments. In theory, with our modulation, detail visibility should be preserved when brightness is reduced, as opposed to substantial contrast loss when constant-dimming.

\paragraph{Study}
A total of 17 people naive to the study with normal vision participated in the experiment (ages 23-31). For each of the four video described above, the two alternatives \textsc{ours} and \textsc{baseline} are shown in a randomized order from one subject to the next. The task of the participants was to give a score 1-5 to rate the quality and saliency of the details found in the scene for each alternative, where 1 indicates low quality and 5 indicates high quality.

\paragraph{Results}
Table \ref{tab:quality-exp} presents the results obtained for all four sequences. Our method consistently outperformed the baseline, receiving higher ratings for perceived detail quality in both scenarios. %

\subsection{Contrast Preservation through a Performance Experiment}
In order to quantitatively measure the preservation of contrast provided by our approach in the previous sequences, we designed a task performance experiment based on finding various Landolt rings \cite{colenbrander1988visual} (in the \textsc{school} and \textsc{house} scenes). We select specific keypoints in these sequences, locating the user in these positions, and maintain the predicted brightness modulation from the previous experiment. The objective is to quantitatively measure (via detection time) if our brightness modulation technique enhances the visibility of environmental details compared to the baseline method. We include extensive details on the stimuli and study details in the Supplemental.

\paragraph{Results}
Figure \ref{fig:performance_comparison} presents the average detection times for each location, along with the standard error of the mean. We can observe that detection times were significantly reduced in low-brightness environments. Furthermore, in bright areas, our modulation did not significantly impact detection time, while saving energy with respect to the baseline via reduced brightness. In rare cases (e.g. \textit{Corridor}), however, our method reduces brightness too aggressively to maintain effective contrast. Although our algorithm achieves consistent contrast losses, it operates globally, which may lead to specific features being compromised, such as the Landolt in this case. This is further discussed in Section \ref{sec:limitations}.

\begin{figure}
    \centering
    \includegraphics[width=\columnwidth]{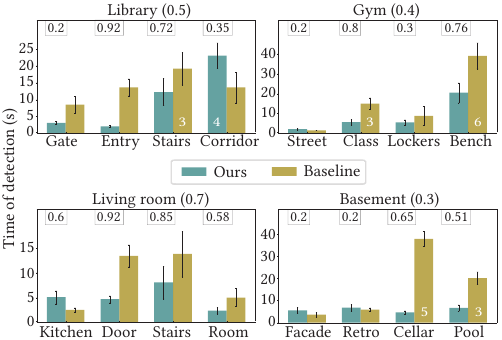}
    \caption{For each location in the four sequences, we report the average detection time. The white numbers indicate trials where the ring was undetected, and the number above each column shows the modulated brightness used. The baseline brightness, in brackets, is constant across all locations in each sequence.}
    \label{fig:performance_comparison}
\end{figure}

\subsection{Imperceptible Brightness Modulation}
\label{sec:brightness-exp}
We also verify that our adjustments in brightness are subtle or imperceptible. While our method primarily focuses on preserving spatial contrast, ensuring temporal coherence is important to prevent artifacts resulting from excessively steep brightness modulation rates. The stimuli utilized in this subjective experiment were identical to those described in Section \ref{sec:quality}.

\paragraph{Study}
The same 17 participants who took part in the detail preservation experiment also participated in this study. Their task was to identify, using a 2AFC protocol, the stimulus that they believed showed a change in brightness. The two alternatives were \textsc{ours} (modulated) and \textsc{baseline} (constant dimming) for each sequence examined. %

\paragraph{Results}
The probability of selecting our method was reported to be nearly 50\% for all sequences (Table \ref{tab:quality-exp}, last column), indicating no substantial perceptual difference between \textsc{ours} and the \textsc{baseline}. This suggests that our method achieves a brightness modulation rate that is imperceptible.

\begin{table}
    \centering
    \caption{Results of both the detail quality experiment and the brightness modulation rate experiment (Sections \ref{sec:quality} and \ref{sec:brightness-exp}, respectively). The first two columns show the comparison of quality ratings for perceived details between \textsc{ours} and the \textsc{baseline}, along with standard error of the mean. 
    Higher values indicate better perceived details.
    The last columns is the probability of detecting brightness changes in the sequence where our algorithm, modulating the brightness, is applied. A result of 0.5 indicates imperceptible modulation.}
    \begin{tabular}{lcc|c}
    \toprule
    \textbf{Sequence} & \textbf{Baseline} & \textbf{Ours} & \textbf{\% 2AFC artifacts det.} \\
    \midrule
        Library & $3.59 \pm 0.22$  & $3.76 \pm 0.16$ & 53\%\\
        Gym & $3.23 \pm 0.26$ & $3.71 \pm 0.19$ & 47\%\\
        Living room & $3.82 \pm 0.19$ & $4.18 \pm 0.17$ & 47\%\\
        Basement & $3.47 \pm 0.15$ & $4.29 \pm 0.18$ & 53\%\\
    \bottomrule
    \end{tabular}
    \label{tab:quality-exp}
\end{table}

\subsection{Ablation on Visual Adaptation}
\label{sec:visualadapt}
\begin{figure}
    \centering
    \includegraphics[width=\columnwidth]{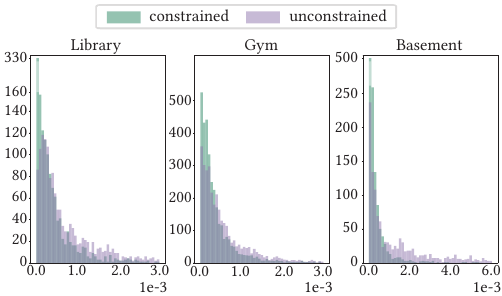}
    \caption{Histogram of the differences between static and dynamic cone responses. Under constrained conditions, the majority of differences are minimal, whereas the unconstrained approach results in larger discrepancies. This demonstrates that our calibration is crucial for maintaining effective visual adaptation throughout the sequences.}
    \label{fig:adaptation}
\end{figure}

By modeling the temporal behavior of photoreceptors, simulating luminance adaptation, we verify that our modulation aligns with the continuous adaptation process of the human visual system. Using Pattanaik's method \shortcite{pattanaik2000time}, we model both the fast, symmetric neural adaptation effects and the slower, asymmetric effects resulting from pigment bleaching, regeneration, and saturation. We compute the static response \cite{hunt2005reproduction}, which assumes viewers have achieved a steady-state of adaptation, and the dynamic time-varying response, that eventually reaches the same state given enough time, indicating full luminance adaptation. For our analysis, we focus exclusively on cones, which are primarily responsible for visual perception when viewing displays under typical daylight conditions. We integrate the values of the cone responses over time and report the results in Table \ref{tab:adaptation}. Furthermore, we show the histogram of differences between static and dynamic cone responses in Figure~\ref{fig:adaptation}.

As we can see, in the tested sequences, the two responses closely align with each other when applying our modulation rate constraint. In contrast, optimizing without this constraint results in a larger divergence between the two responses. This demonstrates that our calibration helps maintaining a continuous visual adaptation of the human eye while watching the displayed content. %

\begin{table}
    \caption{For each sequence, we report the average difference between the static and dynamic cone responses. Results confirm that our modulations %
    closely align with the adaptation process of the HVS. %
    }
    \centering
    \begin{tabular}{lcccc}
    \toprule
    \textbf{Sequence} & \textbf{Brightness factor} & \textbf{Constr.} & \textbf{Unconstr.} \\ 
    \midrule
    Living room & 0.7 & 0.6055 & 0.6151 \\
    Library & 0.5 & 0.4189 & 0.7311 \\
    Gym & 0.4 & 0.4537 & 0.6286 \\
    Basement & 0.3 & 0.3368 & 1.4984 \\
    \bottomrule
    \end{tabular}
    \label{tab:adaptation}
\end{table}

%% file: 5-power.tex
\section{Measuring Real-life Power Savings}
\label{sec:hardware_raspi}

Isolating the power consumption of individual elements in a real VR setup can prove difficult without direct access to the hardware. In order to realistically assess the potential power savings of our approach, and in a similar fashion to closely related works~\cite{cheng2004power}, we crafted a representative setup consisting of an off-the-shelf LCD panel (see Supplemental for the technical sheet), a custom screen driver and a power measuring circuit. A detailed schematic can be seen in the Supplemental. 
To characterize the energy consumption of the display, we measured power consumption at different brightness levels and fit the measurements to establish the relationship between luminance and power consumption, resulting in the following model:

\begin{equation}
    \mathcal{P}(L) = 0.001858 L + 0.2945
\end{equation}
where $L$ is the luminance value; a linear correlation between backlight brightness and power consumption is typical in global-backlight LCD displays. Thanks to the linear correlation, brightness plots throughout our paper can also provide sufficient information on instantaneous power consumption (e.g. Figure~\ref{fig:brightness-contrast}). This function is then used to compute power consumption per frame, $\mathcal{P}_i$, in our optimization (Equation \ref{eq:contrast_minimization}). 

In Table~\ref{tab:powers}, we report measured average screen power consumption throughout Section~\ref{sec:quality}'s experiment. We also report average power consumption when running the sequences without dimming the display (full brightness). Our approach achieves the desired power consumption target while substantially improving perceived quality, as demonstrated before. %

\begin{table}[ht]
    \centering
    \caption{Average screen power consumption throughout the duration of the sequences running on our setup. As imposed by the optimization, \textsc{ours} and \textsc{baseline} consume the same amount of power.}
    \begin{tabular}{lccc}
    \toprule
    \textbf{Sequence} & \textbf{Baseline} & \textbf{Ours} & \textbf{Full brightness} \\
    \midrule
        Library & 0.959 W  & 0.975 W & 1.800 W \\
        Gym & 0.800 W & 0.793 W & 1.797 W \\
        Living room & 1.258 W & 1.259 W & 1.789 W \\
        Basement & 0.646 W & 0.644 W & 1.789 W \\
    \bottomrule
    \end{tabular}
    \label{tab:powers}
\end{table}

%% file: 6-real-time.tex
\section{Extension to Real-Time rendering}
\label{sec:realtime}
While optimal perceived-quality-aware brightness modulation requires knowledge over future content to be displayed, we can still leverage our contrast-luminance sensitivity model to diminish the perceived impact on quality loss when attempting to reduce screen power consumption. Inspired by classic control theory, we developed an alternative formulation of our algorithm that dynamically adjusts screen brightness based on a user-defined maximum tolerable contrast loss. We implemented a proof of concept on GPU with HLSL and Compute Shaders in Unity, which runs real-time.

\begin{figure}
    \centering
    \includegraphics[width=\columnwidth]{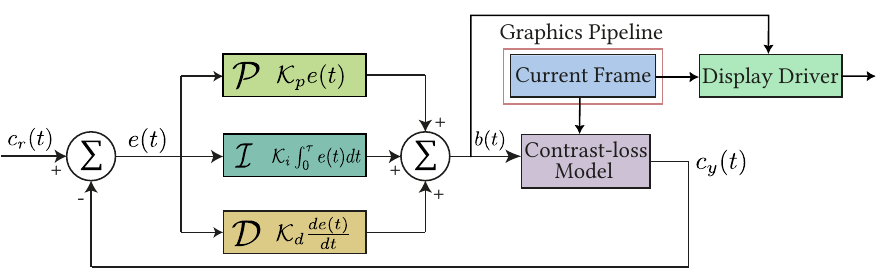}
    \caption{Block diagram of our PID-based control scheme. $c_r(t)$ is the user-defined reference contrast loss, which could be fixed or changed over time. The controller's duty is to achieve this contrast loss by minimizing the error $e(t)$ between the current contrast loss $c_y(t)$ and the desired one. The resulting action from our controller (brightness, $b(t)$) is directly applied to the display driver and is used as well to predict the contrast loss induced by the current displayed frame using the model introduced in Section~\ref{sec:contrast-loss}. We empirically found $\mathcal{K}_p = 0.2$, $\mathcal{K}_i = 0.01$, $\mathcal{K}_d = 0.05$ to work well for a variety of content.}
    \label{fig:pid}
\end{figure}

\paragraph{Method Motivation}
A key insight from our approach in Section~\ref{method} is that it can be seen as closed-loop negative feedback system, where measured contrast loss between subsequent frames influences the brightness of the following frame, attempting to rectify deviations. This conception of the system allows us to leverage classic control theory methods to design optimal real-time brightness modulation. In particular, we deploy a Proportional-Integral-Derivative (PID) controller (Figure \ref{fig:pid}) in conjunction with our previously introduced perceived luminance-contrast sensitivity model, managing average contrast loss to hover over desired values. While directly controlling power consumption would be more desirable, this is unfeasible within our framework in an online/real-time environment, where how the player interacts with the content is outside of our control. In practice, our model saves energy (on average) while maintaining consistent quality; a straightforward alternative (i.e. constant dimming of the screen) attempting to obtain similar energy savings results in degraded quality depending on displayed content (which we specifically test in the \textit{Results} section below).

\paragraph{Implementation details}
Each rendered frame is first processed with the same pipeline described in Section \ref{method}. The algorithm computes the band decomposition and from it the loss of visible contrast with respect to the previous frame. The controller operates with a predefined target contrast loss, adjusting brightness to achieve this target. A higher target contrast loss generally enables more aggressive dimming. With regards to the controller, we can see an overview in Figure~\ref{fig:pid}. The PID's constants $\mathcal{K}_p$, $\mathcal{K}_i$, $\mathcal{K}_d$ were empirically determined across varied content to ensure stability and quick reaction time while avoiding large, sudden brightness changes. 
Broadly, the proportional component $\mathcal{K}_p$ regulates the reaction time of the system to changes in the state: larger values will typically allow it to quickly rectify deviations in the target contrast loss; but too large and it can produce significant overshoots, temporal artifacts and even slower response times. The derivative component $\mathcal{K}_d$ on the other hand, regulates the instantaneous modulation rate and allows us to constrain it in a similar fashion to our temporal calibration in Section~\ref{sec:calibration}. The integral component $\mathcal{K}_i$ smooths the response ensuring that on average we reach our quality target. We include the values we employed throughout our experiments in Figure~\ref{fig:pid}, and speculate about optimal control schemes in Section~\ref{sec:limitations}.

\begin{figure}
    \centering
    \includegraphics[width=\columnwidth]{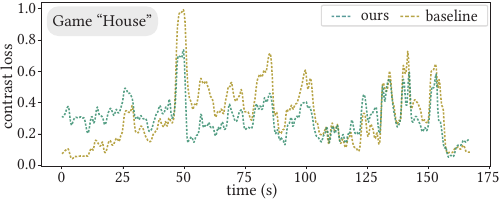}
    \caption{The loss of contrast during the execution of about three minutes of free gameplay in the House environment. The PID controller generates a curve more stable than the baseline ($\sigma_{ours} = 0.12$ and $\sigma_{baseline} = 0.19$).}
    \label{fig:house-wild}
\end{figure}

\begin{wrapfigure}{r}{0.5\columnwidth}
    \centering
    \includegraphics[width=0.48\columnwidth]{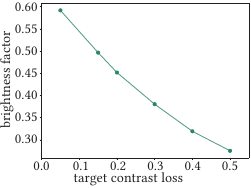}
    \caption{The relationship between target contrast loss and brightness factor.}
    \label{fig:CL-BF}
\end{wrapfigure}

\paragraph{Results}
To verify that visible contrast is preserved, we run a game session of about three minutes in our environment \textsc{House} while using our online approach. We computed average brightness and compared it to a baseline method using it as a constant dimming factor (Figure \ref{fig:house-wild}). Our control scheme ensures that the loss of contrast is more stable throughout the session, improving perceived quality and user experience.

We also study the impact in practical power savings of different contrast loss targets. We run the \textit{Basement} sequence with our real-time approach setting increasingly higher contrast loss tolerances (Figure \ref{fig:CL-BF}). As expected, higher target contrast losses result in substantially lower power usage. The question of where should this threshold be for losses to be imperceptible or acceptable, however, remains as a fascinating avenue for future work.

In terms of performance, our implementation in Unity has an overhead of around $\sim{}25$ms on a computer with an Intel Core i7-9700K and an NVIDIA RTX 2080. It should be possible however to drastically reduce this overhead by further optimizing the GPU implementation, as shown in Tariq et al.~\shortcite{tariq2023perceptually}, which uses a similar luminance-contrast model.

%% file: 7-limitations.tex
\section{Limitations and future work}
\label{sec:limitations}
The effectiveness of our method depends on the content. Sequences with little brightness and detail variation will benefit less from our approach. An interesting extension to our method would involve guiding the optimization with additional information about important regions within the sequence. This would enable prioritizing contrast preservation in specific frames while enabling more aggressive dimming in frames without crucial content. Our method can also be combined with contrast enhancement techniques. We did not explore this in our work and view these methods as independent, leaving the combination for future exploration.

The online alternative of our method has not been perceptually validated with subjective experiments involving participants. While promising as a proof of concept, a more careful controller design could substantially improve its behavior and runtime performance. For example, there are many different approaches, e.g., \cite{evans1950control}, to define optimal control schemes based on a set of constraints (i.e., overshoot, peak transient response time). The challenge in applying them is their requirement for a robust mathematical model of the system. Alternatively, natural image statistics could be leveraged to learn the optimal parametrization for our PID-controlled system.

Our experiments used a power consumption model derived for an LCD panel with a uniform backlight. However, this model is not applicable to all display technologies. Our method can still be used across various technologies; however, the power consumption model should be adjusted, and the exact amount of savings will vary depending on the specific technology. An interesting direction for future work is to extend our method to exploit the local dimming capabilities of displays.

Another exciting venue for future research is exploring saccadic suppression. Similar to other perceptual techniques \cite{sun2018,groth2024} that utilize saccadic eye movements or blinks to hide content manipulation, it may be possible to perform quicker updates to display brightness during such events. This could enable more agressive dimming and, as a result, greater power savings.

%% file: 8-conclusion.tex
\section{Conclusions}
\label{conclusions}
Power efficiency presents a significant challenge for emerging VR technologies, which demand increasingly higher image quality and performance. In this paper, we have introduced a novel technique for managing image brightness over time to achieve a desired power budget while maintaining uniform temporal contrast loss. Our work enhances detail visibility and perceived quality without introducing perceivable artifacts. Furthermore, we have extensively validated our approach, which has been shown to align with visual adaptation models, through various perceptual experiments and corroborated our power saving claims through a custom hardware proxy. Additionally, by tracing connections to control theory, we have introduced a proof-of-concept real-time formulation of our approach, achieving very promising results and increasing the applicability of our method. We believe our framework could spark further interest in adaptive brightness modulation featuring more robust low-level vision models, and it could be of interest to varying fields ranging from video streaming and VR cinema to VR games.

%% file: 9-figures.tex
\newpage
\clearpage

\begin{figure*}
    \centering
    \includegraphics[width=\textwidth]{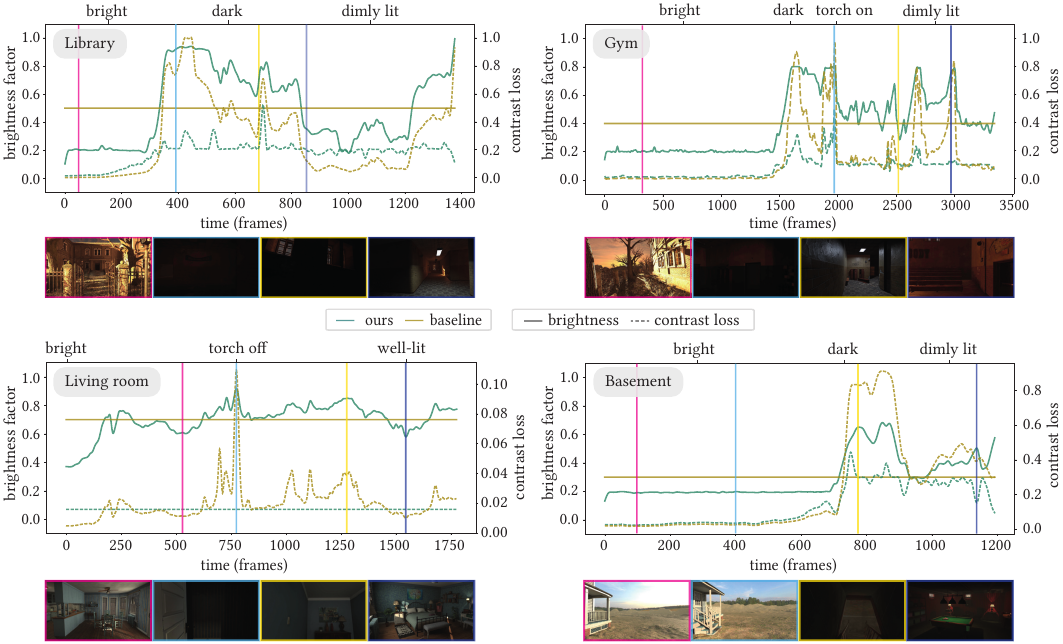}
    \caption{Brightness and loss of contrast plots for the four sequences \textsc{library}, \textsc{gym}, \textsc{living room} and \textsc{basement}. %
    The sequences target an average power consumption a fraction of the maximum screen power (0.6 in \textsc{library}, 0.4 in \textsc{gym}, 0.7 in \textsc{living room} and 0.3 in \textsc{basement}%
    As an example, in the \textsc{library} sequence, the brightness initially starts at a low level and gradually increases when entering indoor spaces with dimmer lighting. Subsequently, the camera navigates through a corridor with moderate illumination before reaching the dark library shelves. Although the modulation rate appears quite steep, it remains within the limit imposed by the calibration (Section \ref{sec:calibration}). %
    Vertical colored lines indicate the timestamps of some example frames shown below.}
    \label{fig:brightness-contrast}
\end{figure*}